# On the colour of thermal noise in fluids

Jana Tóthová[1], Lukáš Glod[2] & Vladimír Lisý[1]

[1]Department of Physics, Technical University of Košice, Park Komenského 2, 04200 Košice, Slovakia. [2] Department of Mathematics and Physics, University of Security Management, Kukučínova 17, 04001 Košice, Slovakia

In the paper by Franosch *et al.*[1] the positional fluctuations of Brownian microspheres in fluids were studied by confining the particles in an optical trap. Experimental access to short timescales has revealed a resonance peak in the spectrum of these fluctuations, in contrast to the commonly assumed overdamped motion. This work is also interesting as the first measurement of the "colour" of thermal noise driving the Brownian particles through collisions with the fluid molecules. The obtained results are described by the hydrodynamic theory of Brownian motion (BM) in harmonic potentials[2]. The interpretation should be, however, improved. We show that the correlation properties of the thermal noise differ from those determined in Ref.[1].

It is long known[3] that the standard Langevin equation (LE) describing BM of particles is valid only under limited conditions. In a more general theory instead of the Stokes friction the resistance force must reflect the memory in the BM. Assuming incompressible fluids, this is the Boussinesq-Basset force[4]. The LE for the velocity $\upsilon(t) = dx/dt$ of Brownian particles then has the form

$$m\dot{\upsilon}(t) + \gamma\upsilon(t) + \int_{t_0}^{t} \Gamma(t-t')\dot{\upsilon}(t')dt' = F_{\text{th}}(t), \qquad (1)$$

where $\Gamma(t) = \gamma(\tau_f/\pi t)^{1/2}$ with $\tau_f = R^2\rho_f/\eta$, $R$ is the radius of the particle, $\rho_f$ is the fluid density and $\eta$ its viscosity, $\gamma = 6\pi\eta R$ is the Stokes friction coefficient, and $m = m_p + m_f/2$, with $m_f$ being the mass of the fluid displaced by the particle of mass $m_p$. The particle is in thermal equilibrium with the liquid. The time $t_0$ is an initial moment infinitely remote from $t$. When the particle is in an external harmonic field, the force $-Kx(t)$, where $x(t)$ is the particle displacement from the trap centre, should be added in the right side of (1). In the traditional LE the thermal noise force $F_{\text{th}}(t)$ is white. This is not the case here since due to the fluctuation-dissipation theorem[5] at different times the values of $F_{\text{th}}(t)$ correlate. The authors[1] have accessed the correlations in the coloured $F_{\text{th}}(t)$ by recording the positions of the particle and calculating the autocorrelation function $\langle x(t)x(0)\rangle$. It was used that at long times the trapping force dominates over friction. Ignoring the particle inertia, the LE was thus reduced to $Kx(t) \approx F_{\text{th}}(t)$. Then it was assumed that $\langle F_{\text{th}}(t)F_{\text{th}}(0)\rangle \approx K^2\langle x(t)x(0)\rangle$. However, this requires that also $Kx(0) \approx F_{\text{th}}(0)$ holds, which is not true; as opposite, $Kx(t)$ is at $t \to 0$ less important than other terms in the LE. Here we show that the calculation of $\langle F_{\text{th}}(t)F_{\text{th}}(0)\rangle$ is possible without these approximations.

Within the linear theory, the properties of $F_{th}(t)$ do not depend on the external force[5]. To find $\langle F_{th}(t)F_{th}(0)\rangle$, it is thus possible to use the LE without $Kx(t)$ and proceed as follows. One can rewrite (1) for $\upsilon(t_0 + t)$, $t > 0$, transforming the integral to $\int_0^t$, multiply this equation by $F_{th}(t_0) = m\dot{\upsilon}(t_0) + \gamma\upsilon(t_0)$, and statistically average. For stationary processes $\langle\dot{\upsilon}(t_0)\upsilon(t_0+t)\rangle = -\langle\upsilon(t_0)\dot{\upsilon}(t_0+t)\rangle = -\dot{\phi}(t)$ and all the terms in the resulting equation for $\langle F_{th}(t_0)F_{th}(t_0+t)\rangle$ can be expressed through the velocity autocorrelation function $\phi(t)$, e.g. $\langle\dot{\upsilon}(t_0)\dot{\upsilon}(t_0+t)\rangle = (d/dt)\langle\dot{\upsilon}(t_0)\upsilon(t_0+t)\rangle = -\ddot{\phi}(t)$. Then, using $\dot{\phi}(0) = 0$ and $\phi(0) = k_BT/m$ (equipartition), the Laplace-transformed equation for $\tilde{\phi}(s) = \mathcal{L}\{\phi(t)\}$ is found. In the theory of hydrodynamic BM[6] $\tilde{\phi}(s) = k_BT\{\gamma + s[m + \tilde{\Gamma}(s)]\}^{-1}$, so that $\mathcal{L}\{\langle F_{th}(t)F_{th}(0)\rangle\} = k_BT[\gamma + \tilde{\Gamma}(s)(s - 1/\tau)]$, where $\tilde{\Gamma}(s) = \gamma(\tau_f/s)^{1/2}$ and $\tau = m/\gamma$. Consequently, the inverse transform for $t > 0$ reads

$$\langle F_{th}(t)F_{th}(0)\rangle = -k_BT\gamma\left(\frac{\tau_f}{\pi t}\right)^{1/2}\left(\frac{1}{\tau} + \frac{1}{2t}\right). \tag{2}$$

Only the term $\sim t^{-3/2}$ has been found in Ref.[1]. Note that the missed term is longer-lived. Using the solution[2] for $\langle x(t)x(0)\rangle$, one can express the right side of (2) through $\langle x(t)x(0)\rangle$ at $t \gg \tau_f$. The resulting formula that should be used to determine $\langle F_{th}(t)F_{th}(0)\rangle$ from the measured $\langle x(t)x(0)\rangle$ at long times is $\langle F_{th}(t)F_{th}(0)\rangle \approx K^2\langle x(t)x(0)\rangle(1 + 2t/\tau)$.

____________________________________

The above text exactly corresponds to the manuscript submitted to (and rejected by) *Nature* as a Brief Communication Arising. Below we give a few additional notes.

**(1)** Instead of Eq. (1) (with $t_0 = -\infty$), one can use the equation

$$M\dot{\upsilon}(t) + \gamma\upsilon(t) + \int_0^t \Gamma(t-t')\dot{\upsilon}(t')\,dt' = F + \zeta(t), \qquad (3)$$

where $\zeta(t) = F_{\text{th}}(t) - \int_{-\infty}^0 \Gamma(-t')\dot{\upsilon}(t')\,dt'$ is a new random force and $F$ is an external force.

One must take into account that whereas $\langle F_{\text{th}}(t)\upsilon(0)\rangle = 0$ due to causality, now the correct solution can be obtained only assuming that $\langle \zeta(t)\upsilon(0)\rangle = (-k_B T\gamma/M)(\tau_R/\pi t)^{1/2}$, i.e., the force $\zeta(t)$ and the velocity $\upsilon(0)$ correlate at $t > 0$.[7] This is in disagreement with some previous results from the literature[8], where the condition $\langle \zeta(t)\upsilon(0)\rangle = 0$ is considered as a "fundamental hypothesis" for solving the Langevin equation (3). Our result for the correlator $\langle \zeta(t)\upsilon(0)\rangle$ can be proved also coming from the basic theorem of the linear response theory (a generalization of the first fluctuation dissipation theorem)[5]; for details see Ref.[7].

The correlation function for $\zeta(t)$ can be obtained in a similar way as above and it is the same as for the "true" thermal force $F_{\text{th}}(t)$. It is seen from the relation between $F_{\text{th}}(t)$ and $\zeta(t)$ below Eq. (3), due to which $\langle \zeta(t)\zeta(0)\rangle = \langle F_{\text{th}}(t)F_{\text{th}}(0)\rangle + C$. The time-independent term $C$ is zero since at $t \to \infty$ both the correlators must converge to zero.

**(2)** The relation between the positional autocorrelation function $\langle x(t)x(0)\rangle$ and $\langle F_{\text{th}}(t)F_{\text{th}}(0)\rangle$ at long times, $\langle F_{\text{th}}(t)F_{\text{th}}(0)\rangle \approx K^2\langle x(t)x(0)\rangle(1 + 2t/\tau)$, follows from Eq. (2) and the long-time asymptote for $\langle x(t)x(0)\rangle \approx -k_B T\gamma(\tau_R/4\pi t^3)^{1/2}/K^2$.[2,7]

**(3)** After this work was completed, we discovered an old paper by Widow[9], which supports our calculations. Although neither the correlation function for the thermal noise nor the correlator $\langle \zeta(t)\upsilon(0)\rangle$ are given in Ref.[9], Eq. (9) in that paper implicitly requires the equality $\langle \zeta(t)\upsilon(0)\rangle = (-k_B T\gamma/M)(\tau_R/\pi t)^{1/2}$, exactly as discussed above in (1).

**(4)** Note that the correlation functions for the thermal noise in incompressible fluids have been calculated also in the recent works[10,11], where the same results are presented as in the commented Ref.[1]

**(5)** The function $\langle F_{\text{th}}(t)F_{\text{th}}(0)\rangle$ is presented in Fig. 1 together with $\langle x(t)x(0)\rangle$ (both the exact result from Eq. (3) with $F = -Kx$ and its long-time asymptote, using the parameters from Ref.[12]: Fig. 2, blue). Figure 2 shows how the initially positive $\langle x(t)x(0)\rangle$ takes negative values and converges to 0 at long times.

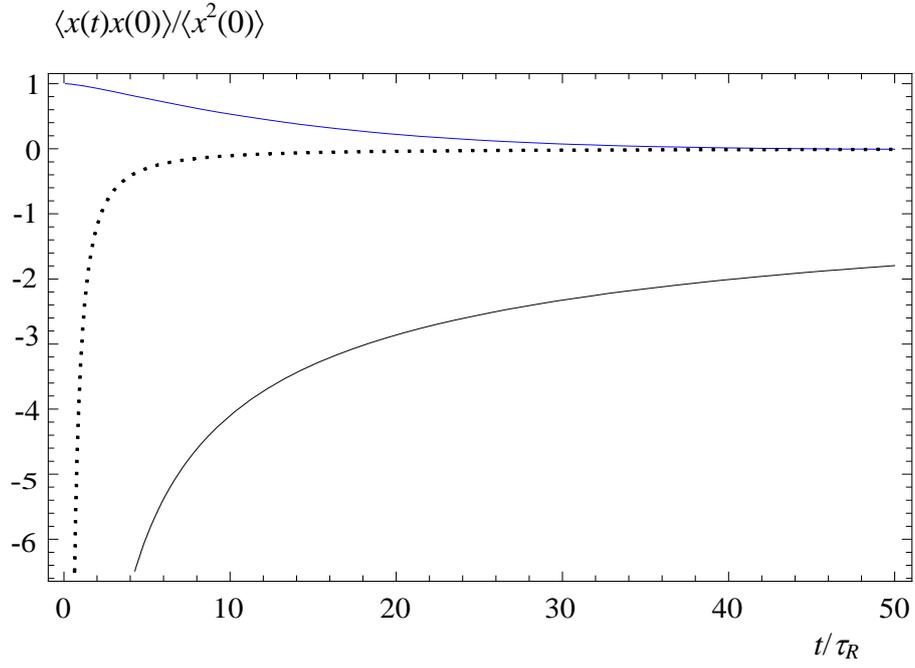

**Fig. 1.** Normalized autocorrelation function $\langle x(t)x(0)\rangle$ calculated from Eq. (3) (upper line) and its long-time limit (points). In contrast to Ref.[1], the correlation function of thermal noise $\langle F_{th}(t)F_{th}(0)\rangle/K^2\langle x^2(0)\rangle$ (lower full line) shows a very different dependence on time.

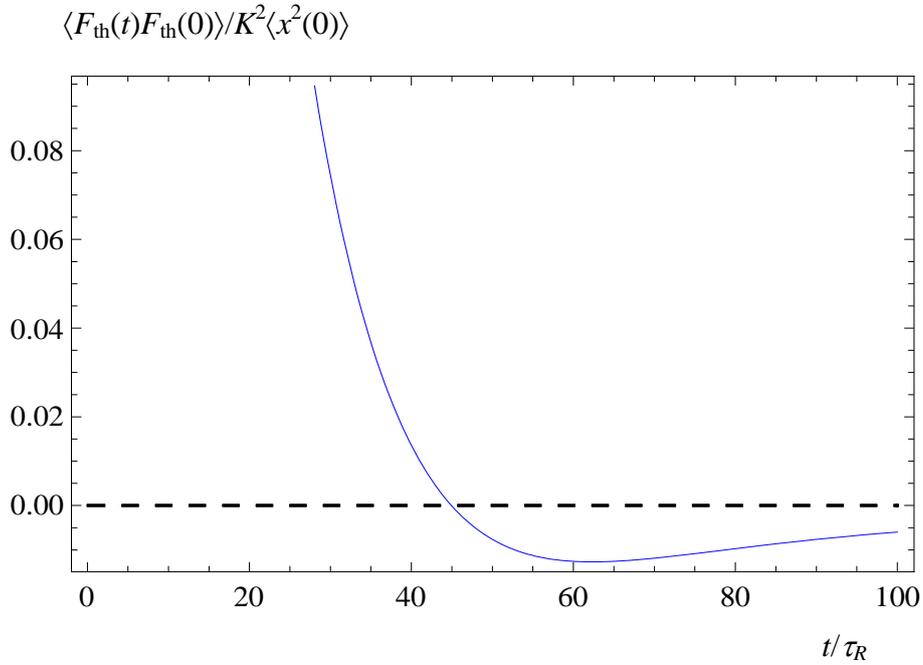

**Fig. 2.** Behaviour of $\langle F_{th}(t)F_{th}(0)\rangle/K^2\langle x^2(0)\rangle$ from Fig. 1 when crossing the zero point.

**ACKNOWLEDGMENT**: This work was supported by the Agency for the Structural Funds of the EU within the projects NFP 26220120021, 26220120033, 26110230061, and by the grant VEGA 1/0370/12.